 \documentclass[final,1p,times]{elsarticle}

\usepackage{amssymb}

\newcommand{\NPA}[3]{Nucl.\ Phys.\ {\bf A#1} (#3) #2}

\newcommand{\PLB}[3]{Phys.\ Lett.\ B\ {\bf #1} (#3) #2}

\newcommand{\PRL}[3]{Phys.\ Rev.\ Lett.\ {\bf #1} (#3) #2}

\newcommand{\PRC}[3]{Phys.\ Rev.\ C\ {\bf #1} (#3) #2}
\newcommand{\PRD}[3]{Phys.\ Rev.\ D\ {\bf #1} (#3) #2}
\newcommand{\JPG}[3]{J.\ Phys.\ G\ {\bf #1} (#3) #2}

\begin{document}

\begin{frontmatter}



\title{Equation of state in hybrid stars and the stability window of quark matter}


\author{Xin-Jian Wen}
\ead{wenxj@sxu.edu.cn}
\address{Department of Physics and Institute of Theoretical Physics,
Shanxi University, Taiyuan 030006, China}

\begin{abstract}
Properties of hybrid stars with a mixed phase composed of asymmetric
nuclear matter and strange quark matter are studied. The quark phase
is investigated by the quark quasiparticle model with a
self-consistent thermodynamic and statistical treatment. We present
the stability windows of the strange quark matter with respect to
the interaction coupling constant versus the bag constant. We find
that the appearance of the quark-hadron mixed phases is associated
with the meta-stable or unstable regions of the pure quark matter
parameters. The mass-radius relation of the hybrid star is dominated
by the equation of state of quark matter rather than nuclear matter.
Due to the appearance of mixed phase, the mass of hybrid star is
reduced to 1.64 M$_{\odot}$ with radius $10.6$ km by comparison with
neutron star.

\end{abstract}


\end{frontmatter}


\section{INTRODUCTION}
\label{sec:intro}

The appearance of quark matter or hadron-quark mixed phase in the
massive neutron stars is a hot topic in the study of compact
objects. The baryon densities of the stars cover a larger range from
very low densities in the outer part to the order of about ten times
the saturation density in the inner core. To study the structure of
compact stars, the key point is to find a reliable form of the
equation of state (EOS) that determines the characteristic of the
constituent matter \cite{Read09}. Unfortunately, however, there is
no single theory to cover the large density range with respect to
quark degrees of freedom. In literature, the low-density phase can
be described by quantum hadrodynamics (QHD). At high densities, a
new form of matter, called strange quark matter (SQM), might exist
and be more stable than nuclear matter ($^{56}$Fe)
\cite{bodmer,witten1984,Liu1984,Crawford1992,Lugones02}. For more
theoretical and experimental results, see
\cite{Berger1987,Madsen1993,Schaffner-Bielich1997,Madsen2001}. A
family of compact stars consisting completely of the deconfined
mixture of $u$-, $d$-, $s$- quarks has been called ``strange stars"
\cite{Alcock1986,Haensel1986,RxXu09}. If the hypothesis of stable
strange quark matter is correct \cite{witten1984}, the possibility
of phase transition exists in principle \cite{Bombaci2008}. The
compact star with a ``quark matter core", either as a hadron-quark
mixed phase or as a pure quark phase, are called ``hybrid stars"
(HyS) \cite{Glendenning1996, Goyal2004}. Recently, it was argued
that the interior core of a low-mass compact star could be dominated
by the Color-Flavor-Locked (CFL) quark matter
\cite{Stejner2005,Alford2003}. In contrast, it was remarked that the
CFL phase would make the HyS unstable \cite{klahn2007} and the
possible configuration of compact stars, such as the strange
hadrons, hyperonic matter \cite{Sahu2001} and quark matter core, can
soften the equation of states of neutron stars
\cite{Sahu2001,Shen2002,Burgio2002}. However, Alford et al pointed
out the ``masquerade effect" \cite{alford2005} that the hybrid star
has a mass-radius relation similar to that of the pure neutron star.
Also, the isolated neutron star RX J1856-375 with a large radius
and/or mass \cite{trumper2004} is a possible candidate of HyS, which
implies a constraint on testing a rather stiff equation of state at
high density \cite{klahn2006}. Regardless of whether the quark
matter is ruled out, it seemed that the soft equation of state were
ruled out in the center of compact stars \cite{ozel2006}.

To get a reliable equation of state in the microscopic calculation
of interacting dense hadronic matter, Wiringa et.al added the
three-body potential to the nucleon Hamiltonian and gave the light
nuclei binding energy and nuclear matter saturation properties
\cite{Wiringa1988}. The Brueckner theory with three-body forces has
been used recently to study the mixed phase of hadrons and quarks in
compact stars \cite{peng2008}. In literature, there is another
successful method called relativistic mean-field (RMF) theory. It is
a powerful tool in describing various aspects of nuclear physics,
such as the properties of nuclear matter, finite nuclei, and neutron
stars, as well as the dynamics of heavy-ion collisions
\cite{Sahu2000,glendenning1987}. Recently, the model has been
extended to include the density-dependent meson-nucleon coupling
constant in finite nuclei \cite{Ring2002} and nuclear matter
\cite{klahn2006,Meng2004}.

In the theoretical description of the deconfined quark matter, we
can resort to the phenomenological models constrained by
experimental information. There are many successful works
considering the medium-effect of quark masses. One of them obtains
the confinement by the density dependence of quark masses
\cite{QMDD}. Another one is the quark quasiparticle model, where the
vacuum energy density is not constant but density-dependent. For the
medium dependence of the quasiparticle masses, it was derived at the
zero-momentum limit of the dispersion relations from an effective
quark propagator by resuming one-loop self-energy diagrams in the
hard-dense-loop approximation (HDL) \cite{Schertler1997}. We have
recently extended the model to include the important finite-size
effect \cite{wen2009} and magnetized strange quark matter (MSQM)
\cite{wen2012}.

In this paper, we study the effect of the interaction coupling
constants on the equation of mixed phase of nuclear and quark matter
and the properties of hybrid stars. For the nuclear EOS, we adopt
the relativistic nuclear field theory solved at the mean-field
level, and especially the Baym-Pethick-Sutherland (BPS) model for
densities below $5\times 10^{14}$g cm$^{-3}$
\cite{Baym1971,ruster2006}. In describing quark matter, we employ
the quark quasiparticle model instead of the conventional bag model
\cite{Burgio2002}. The density-dependent bag function is obtained
self-consistently rather than artificially.

This paper is organized as follow. In Section \ref{Sec:thermo}, we
briefly give a short introduction to the relativistic nonlinear
mean-field model describing the nuclear matter. In Section \ref{QM},
we introduce the treatment of SQM in the framework of the quark
quasiparticle model and present the stability window of quark
matter. In Section \ref{comstar}, we display the phase diagram of
the mixed phase and discuss the chemical potential behavior. With
the equations of state, we investigate the influence of coupling
constant and bag constant on the mass-radius relation. The last
section is a short summary.

\section{The nuclear EOS in the relativistic mean-field model}
\label{Sec:thermo}

The relativistic nuclear field theory is solved at the mean-field
level. The in-medium interaction of nucleons can be realized through
the exchanges of the $\sigma$, $\omega$, and $\rho$ mesons. The bulk
matter is assumed to be electrically neutralized and in the lowest
energy state, i.e. in general $\beta$-equilibrium. The influence of
the temperature can be neglected \cite{Manka2000}. The Lagrange
density for this model is written as \cite{Manka2001,Bednarek2001}
\begin{eqnarray}
{\cal{L}}_{RMF} &=&\frac{1}{2} \left(
\partial_{\mu}\sigma\partial^{\mu}\sigma  +
m_\rho^2\rho_\mu\rho^\mu\right) -U(\sigma)+V(\omega)
 -\frac{1}{4} \left[\Omega_{\mu\nu}\Omega^{\mu\nu}+ R^a_{\mu\nu}R^{a\mu\nu}+F_{\mu\nu}F^{\mu\nu} \right] \nonumber\\
& &+\bar{\psi} \left(i\gamma^{\mu}\partial_{\mu}-m_{N}+g_{\sigma
N}\sigma -g_{\omega N}\gamma^\mu\omega_\mu -\frac{1}{2} g_{\rho
N}\gamma^\mu \rho_\mu^a\cdot \tau^a -e\gamma^\mu Q_e
A_\mu\right)\psi \nonumber\\
&& +\bar{\psi}_e(i\gamma^\mu \partial_\mu-m_e)\psi_e.
\end{eqnarray}
where the nucleon field $\psi$ has a form of column vector for
protons and neutrons, and the field tensors are given by
\begin{eqnarray}
\Omega_{\mu\nu} &=&
 \partial_{\nu}\omega_{\mu}-\partial_{\mu}\omega_{\nu},\\
R^a_{\mu\nu} &=&
 \partial_{\mu}\rho^a_{\nu}-\partial_{\nu}\rho^a_{\mu}, \\
F_{\mu\nu} &=& \partial_{\nu}A_{\mu}-\partial_{\mu}A_{\nu}.
\end{eqnarray}
The non-linear potential functions are contained for the meson
$\sigma$ \cite{Bodmer1991} and $\omega$ as
\begin{eqnarray}
U(\sigma)&=&\frac{1}{2} m_\sigma^2\sigma^2
+\frac{1}{3} g_3\sigma^3 +\frac{1}{4} g_4\sigma^4,\\
V(\omega)&=&\frac{1}{2} m^2_\omega \omega_\mu\omega^\mu+\frac{1}{4}
c_4 (\omega_\mu\omega^\mu)^2.
\end{eqnarray}
The parameters $m_N$, $m_\sigma$, $m_\rho$, $m_\omega$ are the
masses of nucleon, scalar meson $\sigma$, isovector-vector meson
$\rho$, and isoscalar-vector meson $\omega$ respectively. In
principle, they can be fixed algebraically by the properties of bulk
nuclear matter, such as, the binding per nucleon, the saturation
baryon density, the effective mass of the nucleon at saturation, and
the compression modulus. The isovector $\rho$ field vanishes for
symmetry nuclear matter.

The scalar density and conserved baryon number density are expressed
with the use of Fermi integrals
\begin{eqnarray}
\rho_s&=&\frac{4}{(2\pi)^3}\int_0^{p_f}
\frac{M^*}{\sqrt{p^2+M^{*2}}}dp^3,\\
\rho_N&=&\rho_p+\rho_n=\frac{\nu^3_p}{3\pi^2}+\frac{\nu^3_n}{3\pi^2},
\end{eqnarray}
where the effective nucleon mass is defined by $M^*\equiv
M-g_{\sigma N}\sigma_0$, with $\nu_p$ and $\nu_n$ being the Fermi
momenta of protons and neutrons respectively.

The energy density and pressure of nuclear matter can be obtained
from the energy-momentum tensor. Including the contribution of
nucleons and mesons, the total energy density $\epsilon^\mathrm{HP}$
and the pressure $P^\mathrm{HP}$ are
\cite{Glendenning1996,alford2001}
\begin{eqnarray}
\epsilon^\mathrm{HP}&=&\frac{2}{(2\pi)^3}\left[\int_0^{\nu_p}\sqrt{p^2+{M^*}^2}dp^3
+ \int_0^{\nu_n}\sqrt{p^2+{M^*}^2}dp^3 \right]
\nonumber \\
&& +\frac{1}{2} m_\sigma^2\sigma^2 +\frac{1}{3} g_3\sigma^3
+\frac{1}{4} g_4\sigma^4 +\frac{1}{2} m^2_\omega
\omega^2_0+\frac{3}{4} c_4 \omega_0^4 +\frac{1}{2}m^2_\rho
\rho_0^2,\\
P^\mathrm{HP}&=&\frac{1}{3}\frac{2}{(2\pi)^3}\left[\int_0^{\nu_p}\frac{p^2}{\sqrt{p^2+{M^*}^2}}dp^3
+ \int_0^{\nu_n}\frac{p^2}{\sqrt{p^2+{M^*}^2}}dp^3 \right]
\nonumber \\
&& -\frac{1}{2} m_\sigma^2\sigma^2 -\frac{1}{3} g_3\sigma^3
-\frac{1}{4} g_4\sigma^4 +\frac{1}{2} m^2_\omega
\omega^2_0+\frac{1}{4} c_4 \omega_0^4+\frac{1}{2}m^2_\rho
\rho_0^2.\label{pressN}
\end{eqnarray}
If the electron contribution is included,
$P_\mathrm{e}=\mu_e^4/(12\pi^2)$. For symmetric nuclear matter, the
number density of protons and neutrons are equal, and
correspondingly the Fermi momenta are also equal, i.e.,
$\nu_p=\nu_n$. In this case, the isovector meson has no
contribution.

From the equation of motion for nucleons, the Fermi energy of
nucleons, or equivalently, the chemical potential $\mu_N$ can be
expressed as
\begin{eqnarray}
\mu_N=g_{\omega N}\omega_0+\frac{1}{2}g_{\rho
N}\rho_{03}\tau_{3N}+\sqrt{\nu^2+{M^*}^2}.
\end{eqnarray}

In this paper, we choose four typical sets of parameters (TM1, NL3,
BKA20, and TW-99) \cite{Meng2004,BKA}. They stand for different
stiffness equations of state. The messes for $\sigma$, $\omega$, and
$\rho$ mesons are $m_\sigma=509$MeV, $m_\omega=782$MeV, and
$m_\rho=770$MeV with nucleon mass $M=939$MeV. For getting the EOS in
the low-density, i.e., the crust equation of state of compact stars,
we can resort to the BPS model \cite{Baym1971} in calculations.

\section{Quark quasiparticle model for quark matter EOS}
\label{QM}

At high density, nuclear matter is expected to undergo a phase
transition to a deconfined phase. If the quark chemical potential
exceeds the strange quark mass, the system can lower its Fermi
energy by converting the down quark into strange quarks. Recently,
we have developed the quark quasiparticle model in studying
strangelets \cite{wen2009} and MSQM \cite{wen2012}. In the
quasiparticle model, the effective quark mass following from the
hard-dense-loop (HDL) approximation of quark self-energy at zero
temperature is expressed as \cite{Schertler1997,Pisarski1989},
\begin{eqnarray}
m_i^*=\frac{m_i}{2}+ \sqrt{\frac{m^2_i}{4}+
\frac{g^2\mu_i^2}{6\pi^2}}\, ,\label{mass1}
\end{eqnarray}
where $m_i$ is the corresponding current mass of quarks,
$g=\sqrt{4\pi\alpha_s}$ denotes the strong interaction coupling
constant. The effective quark mass $m_i^*$ increases with $g$, $m_i$
and the quark chemical potential $\mu_i$. In this paper, we treat
$g$ as a free parameter in the range of $(0,5)$. For light quarks
($u$ and $d$ quarks), we take the current mass to be zero, and
Eq.~(\ref{mass1}) is reduced to the simple form
\begin{equation}
\label{miud}m_i^*=\frac{g\mu_i}{\sqrt{6}\pi}\equiv\alpha\mu_i, \ \
(i=u, d)
\end{equation}

The quasiparticle contribution to the thermodynamic potential
density is given as \cite{wen2009}.
\begin{eqnarray}
\Omega_i &=&
   -\frac{d_i T}{2\pi^2}
   \int_0^{\infty}
    \left\{
 \ln\left[1+e^{-(\epsilon_{i,p}-\mu_i)/T}\right]
   +\ln\left[1+e^{-(\epsilon_{i,p}+\mu_i)/T}\right]
    \right\}
  p^2 \mbox{d}p,
\end{eqnarray}where
$\epsilon_{i,p}=\sqrt{p^2+{m_i^*}^2}$ and $T$ is the temperature.
At zero temperature, the integration can be calculated out to give
\begin{eqnarray}\Omega_i 
&=&-\frac{d_i}{48\pi^2}
     \Bigg[
|\mu_i|\sqrt{\mu_i^2-{m_i^*}^2}\left(2\mu_i^2-5{m_i^*}^2\right)
  +3{m_i^*}^4\ln\frac{|\mu_i|+\sqrt{\mu_i^2-{m_i^*}^2}}{m_i^*}
    \Bigg],
\end{eqnarray}
 where $m_i$ and $\mu_i$ are, respectively, the particle
masses and chemical potentials. $d_i$ is the degeneracy factor with
$d_i=2(\rm{spin})\times 3(\rm{color})=6$ for quarks and $d_i=2$ for
electrons. When the variable set (T, V, \{$\mu_i$\}) is chosen as
the independent state variables, the thermodynamic potential is the
characteristic function. In the quasiparticle model, the total
thermodynamic potential density can be written as
\begin{equation} \label{Omegtot}
\Omega=\sum_i \left [\Omega_i(\mu_i,m_i^*)+B_i(\mu_i)\right]+B_0,
\end{equation}
Where $B_i(\mu_i)$ is the medium-dependent quantity determined by
thermodynamic consistency. A little later, we will see that
$B_i(\mu_i)$ is given by an indefinite integration. When it is
expressed by a definite integration, an integration constant is
needed. Therefore, $B_0$ is from the sum of relevant integration
constants, and we treat it as a free input parameter. In the
quasiparticle model, the particle number density should be of the
same form as that of a Fermi gas with the normal particle mass
replaced by the effective quasiparticle mass,i.e.,
\begin{eqnarray}\label{nderi}
n_i=-\frac{\partial \Omega_i}{\partial \mu_i}
=\frac{d_i}{6\pi^2}\left(\mu_i^2-{m_i^*}^2\right)^{3/2}.
\end{eqnarray}
On the other hand, we know from the fundamental thermodynamics that
\begin{equation} \label{ni2}
n_i
=-\left.\frac{\mathrm{d}\Omega}{\mathrm{d}\mu_i}\right|_{\mu_{k\neq
i}} =-\frac{\partial\Omega_i}{\partial \mu_i}
 -\frac{\partial \Omega_i}{\partial m_i^*}\frac{\partial m_i^*}{\partial \mu_i}
 -\frac{\mathrm{d} B_i}{\mathrm{d} \mu_i}.
\end{equation}

Equating the last equality in Eq.~(\ref{ni2}) with the first
equality in Eq.~(\ref{nderi}), we immediately have
\begin{equation} \label{Biexp}
\frac{\mathrm{d} B_i}{\mathrm{d} \mu_i} =-\frac{\partial
\Omega_i}{\partial m_i^*}\frac{\partial m_i^*}{\partial \mu_i}
 \ \ \mbox{i.e.} \ \
B_i=-\int_{m_i^*}^{\mu_i}
      \frac{\partial \Omega_i}{\partial m_i^*}
      \frac{\partial m_i^*}{\partial \mu_i}
     \mbox{d}\mu_i,
\end{equation}
where the derivative of the thermodynamic potential density with
respect to the quark effective mass can be analytically expressed by
\begin{eqnarray}\frac{\partial \Omega_i}{\partial m_i^*}&=&\frac{d_i
m_i^*}{4\pi^2} \left[\mu_i \sqrt{\mu_i^2-{m_i^*}^2}  -{m_i^*}^2 \ln
\frac{\mu_i+\sqrt{\mu_i^2-{m_i^*}^2}}{m_i^*}
\right]. \label{dodm}
\end{eqnarray}

Accordingly, the pressure $P^\mathrm{QP}$, energy density
$\epsilon^\mathrm{QP}$, and baryon density $\rho^{QP}$ for SQM at
zero temperature are written as,
\begin{eqnarray}P^\mathrm{QP}&=&-\sum_i \left [\Omega_i(\mu_i,m_i^*)+B_i(\mu_i)\right]-B_0,\label{pressQ}\\
\epsilon^\mathrm{QP} &=& \sum_i \left [\Omega_i(\mu_i,m_i^*)+B_i(\mu_i)\right]+\sum_i \mu_i n_i+B_0, \label{energy}\\
\rho^\mathrm{QP}&=&\frac{1}{3}\sum_i n_i.
\end{eqnarray}

Because the current mass of light quarks is nearly zero, one can
have an analytical expression for light quarks by combining Eqs.\
(\ref{miud}) and (\ref{Biexp}) \cite{wen2009,Schertler1997}, giving
\begin{eqnarray}\label{Bexp}
B_i^*(\mu_i)&=&-\int^{\mu_i}_0 \left.\frac{\partial \Omega}{\partial
m^*_i}\right|_{T=0,\mu_i}\frac{\mathrm{d} m^*_i}{\mathrm{d}\mu_i} \mathrm{d}\mu_i \nonumber \\
&=&-\frac{d_i}{16\pi^2} \left[\alpha^2\beta-\alpha^4
\ln(\frac{\beta+1}{\alpha}) \right]\mu_i^4, \ \ \ (i=u,d),
\end{eqnarray}
where $\alpha=g/(\sqrt{6}\pi)$ and $\beta=\sqrt{1-\alpha(g)^2}$ are
$g$-dependent functions.

For massive quarks, $B^*(\mu)$ has been observed as the result
\cite{Schertler1997},
\begin{eqnarray}
B_s^*(\mu_s)&=&-\frac{d_s}{16\pi^2}\Big[
\frac{\sqrt{(m_{s}^*-m_s)(\beta^2 m_s^*-m_s)}}{24 \alpha^2\beta^4}
\Big] \times \sum_{n=0}^3 a_n m_s^{3-n}{m_s^*}^n -{m_s^*}^4\ln\big( \frac{k_F+\mu}{m_s^*} \big) \nonumber\\
&&  +\frac{5 \alpha^4-12 \alpha^2+8}{16 \beta^5}m_s^4 \times\ln\Big(
\frac{\beta\sqrt{m_s^*-m_s}+\sqrt{\beta^2m_s^*-m_s} }{\alpha^2m_s}
\Big),
\end{eqnarray}
where the coefficient $a_n$ can be related to the coupling constant
through four polynomial function as in \cite{Schertler1997}, which
is very different from the gaussian parametrization of the
density-dependent bag constant \cite{maieron2004}.

\begin{figure}[htb]
\centering
\includegraphics[width=7.cm,height=7cm]{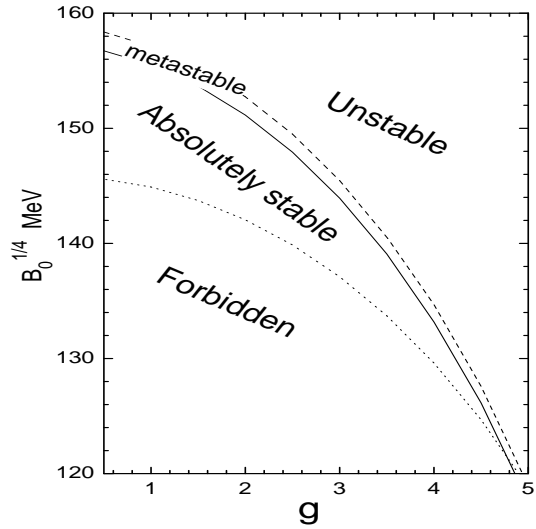}
\caption{Stability windows in the parameter space (B$_0^{1/4}$,
$g$). Below the dotted line is the forbidden area where two-flavor
quark matter is absolutely stable. Above the solid or dashed line,
SQM is meta-stable or unstable, and SQM is absolutely stable between
the the dotted and solid curves. } \label{fig B0window}
\end{figure}

With the above quark mass formulas and thermodynamic treatment, one
can get the properties of bulk quark matter. According to the
Witten-Bodmer hypothesis, it is required that the energy per baryon
of two-flavor quark matter is bigger than 930 MeV in order not to
contradict with standard nuclear physics, but that of symmetric
three-flavor quark matter is less than 930 MeV. For the co-existence
of nuclear matter and quark matter in the process of phase
transition, the allowed values of model parameters should be chosen
in the metastable or unstable regions. In Fig. \ref{fig B0window},
we present the stability windows of SQM in the $B_0^{1/4}$-$g$
plane. The area below the dotted line is forbidden because the
energy per baryon $\epsilon^{QP} /\rho^{QP}$ of 2-flavor quark
matter is less than $930$ MeV. The $\epsilon^{QP} /\rho^{QP}$ of
3-flavor quark matter is smaller than $930$ MeV below the solid line
and is smaller than 939 MeV in the narrow area below the dashed
line. In the top area the strange quark matter is unstable.

\section{Hadron-quark mixed phase and hybrid stars}
\label{comstar}
\subsection{Transition from nuclear matter to quark matter}

We suppose that the compact star composed of quark matter core and
nuclear matter surface. The nuclear matter includes protons,
neutrons and electrons. The quark matter consists of a mixture of
quarks ($u$, $d$, and $s$) and electrons. In the interface of the
two phases, there may be a mixed phase of nuclear and quark matter,
which is important in understanding the properties of hybrid stars.
To describe the structure of the mixed phase, we define the fraction
$\chi$ occupied by quark matter by $\chi\equiv
V^\mathrm{QP}/(V^\mathrm{QP}+V^\mathrm{HP}).$ Then we have the total
baryon number density $\rho$, energy density $\epsilon$, and
electrical charge $Q$ as,
\begin{eqnarray}
\rho_B&=&(1-\chi)\rho^\mathrm{HP}+\chi \rho^\mathrm{QP},\label{baryoneq}\\
\epsilon&=&(1-\chi)\epsilon^\mathrm{HP}+\chi\epsilon^\mathrm{QP},\\
Q&=&(1-\chi)Q^\mathrm{HP}+\chi Q^\mathrm{QP},\label{chargeq}
\end{eqnarray}where $Q=0$ according to the bulk charge neutrality
requirement. The critical density $\rho_c^\mathrm{HP}$ for pure
nuclear matter is determined by $\chi=0$. With increasing density,
the deconfinement phase transition takes place. Consequently, quark
matter appears in the mixed phase. When $\chi$ increases up to $1$,
the quark phase dominates the system completely at the critical
density $\rho_c^\mathrm{QP}$.

From the quark constituent of nucleons, we have the chemical
potential relations through the linear combinations,
\begin{equation} \label{mupnexp}
 \mu_p= 2\mu_u +\mu_d, \ \
 \mu_n= \mu_u+2\mu_d.
\end{equation}
Incorporated with the $\beta$-equilibrium condition
$\mu_d=\mu_s=\mu_u+\mu_e$, there are only two independent chemical
potentials, e.g.\ $\mu_u$ and $\mu_e$. The Gibbs condition for the
phase equilibrium between nuclear matter and quark matter is that
the system should be in thermal, chemical, and mechanical
equilibrium. Therefore, in addition to the common zero temperature,
we also have the two conditions
\begin{eqnarray}
\mu^\mathrm{HP}&=&\mu^\mathrm{QP}, \label{baronpot}
\\
P^\mathrm{HP}(\mu_p,\mu_n,\mu_e)&=&P^\mathrm{QP}(\mu_u,\mu_d,\mu_s,\mu_e),
\label{presseq}
\end{eqnarray}
where $P^\mathrm{HP}$ and $P^\mathrm{QP}$ denotes the pressure in
the nuclear phase and in the quark phase, respectively. They are
given by Eqs. (\ref{pressN}) and (\ref{pressQ}) separately. The
electron is uniformly distributed in the system and the pressure is
common for both phases. Eq.\ (\ref{baronpot}) means that the baryon
chemical potential in both phases are the same. It can be shown that
Eq.\ (\ref{baronpot}) is equivalnet to Eq.\ (\ref{mupnexp})
\cite{peng2008}. For a given total baryon number density, one can
obtain the two independent chemical potentials and $\chi$ by solving
the equations (\ref{baryoneq}), (\ref{chargeq}) and (\ref{presseq}).

\begin{figure}[ht]
\begin{minipage}[t]{0.48\linewidth}
\centering
\includegraphics[width=7.cm,height=7cm]{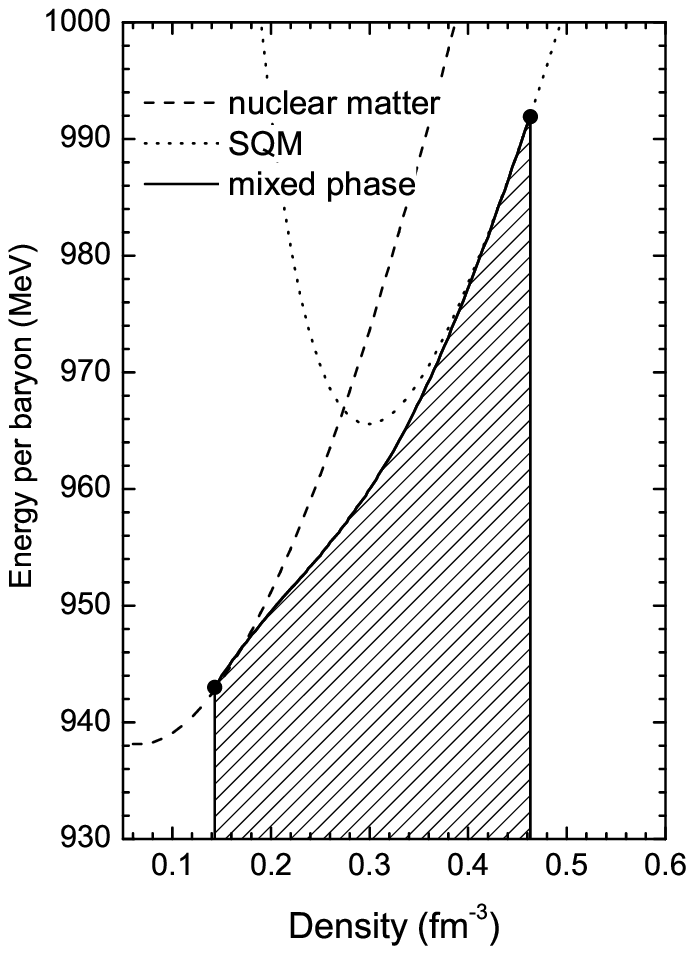}
\caption{The energy per baryon as a function of density. In the
density range of shaded area, the energy per baryon
 (solid line) is lower than that of either quark matter (dotted line)
or nuclear matter (dashed line). $g=3$, $B_0^{1/4}=150$ MeV, and TM1
set are adopted. }\label{fig_energ}
\end{minipage}
\hfill
\begin{minipage}[t]{0.48\linewidth}
\centering
\includegraphics[width=7.cm,height=7cm]{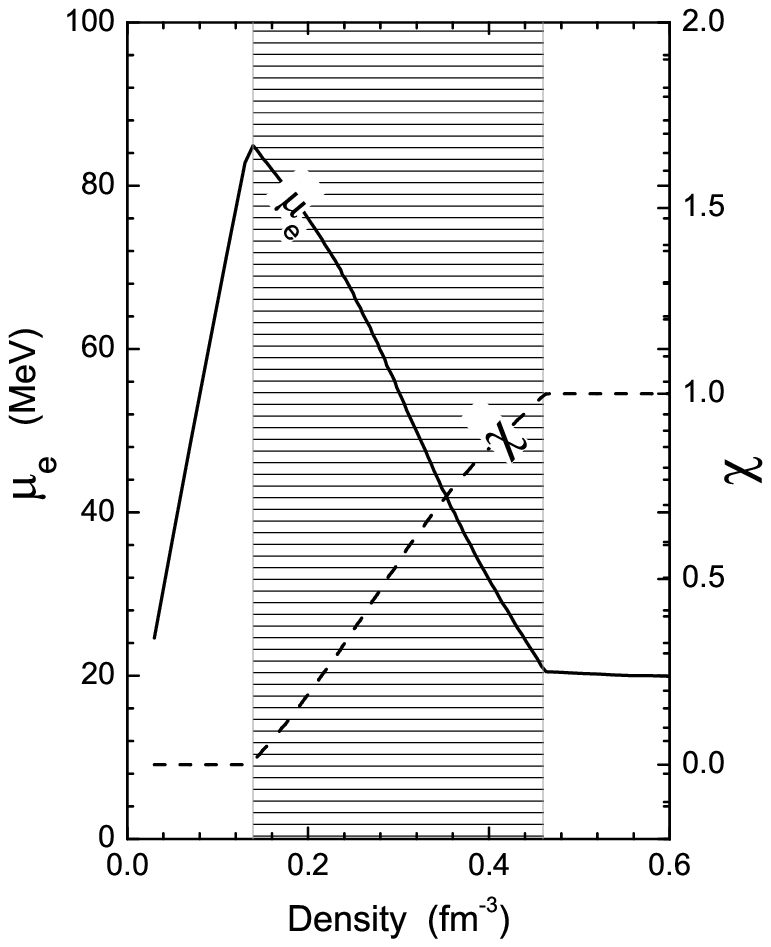}
\caption{ The electron chemical potential $\mu_e$ (left axis) and
the fraction of SQM $\chi$ (right axis). Parameters are the same as
for Fig.\ \ref{fig_energ}.} \label{fig muechi}
\end{minipage}
\end{figure}
In Fig. \ref{fig_energ}, the system energy per baryon is displayed
as a function of the density for $g=3$, $B_0^{1/4}=150$ MeV, and TM1
parameter set. The mixed phase (shaded area) exists in the range of
about $1\sim 3$ times the nuclear saturation density. In this
density range, the energy per baryon of mixed phase (solid line) is
lower than that of both the pure nuclear (dashed line) and the pure
quark phase (dotted line). So the appearance of the mixed phase in
neutron star matter is energetically favored for a proper parameter
in meta-stable or unstable regions. This observation is consistent
with that in Ref. \cite{peng2008}. If one chooses the absolutely
stable parameter-sets ($g$, $B_0^{1/4}$), no mixed phase will exist
and the hybrid star is unstable and will collapse to a strange star.
In the middle density range, the mixed phase starts at the nuclear
critical density $\rho_c^\mathrm{HP}$ where $\chi=0$ and ends at the
quark critical density $\rho_c^\mathrm{QP}$ where $\chi=1$. The
critical points are marked with full dots in Fig. \ref{fig_energ}.
By comparing the different equation of state for nuclear matter, we
find that the solid lines between the $\rho_c^\mathrm{HP}$ and
$\rho_c^\mathrm{QP}$ is shorter for harder EOS of nuclear matter. It
is also observed that the quark critical density
$\rho_c^\mathrm{QP}$ varies only slightly with the RMF parameters.

To provide a better understanding of the Glendenning hypothesis of
global charge neutrality, we give the electron chemical potential
and the quark fraction versus the total density in Fig.\ \ref{fig
muechi}. In the shaded area of mixed phase, the increasing
contribution of quark phase is a complementarity to the decrease of
electrons. The electron chemical potential $\mu_e$ (dashed line)
decreases rapidly from the maximum value $85$ MeV to $20$ MeV. And
finally it becomes very small in the pure quark phase where the
quark fraction $\chi$ is unity, where the three-flavor quark matter
occupies the whole space.

\subsection{The structure of hybrid stars}
\label{Sec:numerical}

With the above equation of state, we now study the structure of
compact stars. As usually done, we assume that the hybrid star is a
spherically symmetric distribution of mass in hydrostatic
equilibrium. The equilibrium configurations are obtained by solving
the Tolman-Oppenheimer-Volkoff (TOV) equation for the pressure
$P(r)$, the energy density $\epsilon(r)$ and the enclosed mass
$m(r)$:
\begin{equation}
\frac{dP(r)}{dr} =-\frac{Gm(r)\epsilon(r)}{r^2}
\frac{[1+P(r)/\epsilon(r)][1+4\pi r^3 P(r)/m(r)]}{1-2Gm(r)/r}
\end{equation} where $G=6.707\times 10^{-45}$MeV$^{-2}$ is the
gravitational constant, $r$ is the distance from the center of the
star. The subsidiary condition is
\begin{equation}\frac{dm(r)}{dr}=4\pi r^2 \epsilon(r).
\end{equation}
Giving the stellar radius $R$, which is defined by zero pressure at
the stellar surface, the gravitational mass is given by
\begin{equation}
M(R)\equiv 4\pi\int^R _0r^2 \epsilon(r)\mbox{d}r.
\end{equation}

\begin{figure}[ht]
\begin{minipage}[t]{0.48\linewidth}
\centering
\includegraphics[width=7.cm,height=7cm]{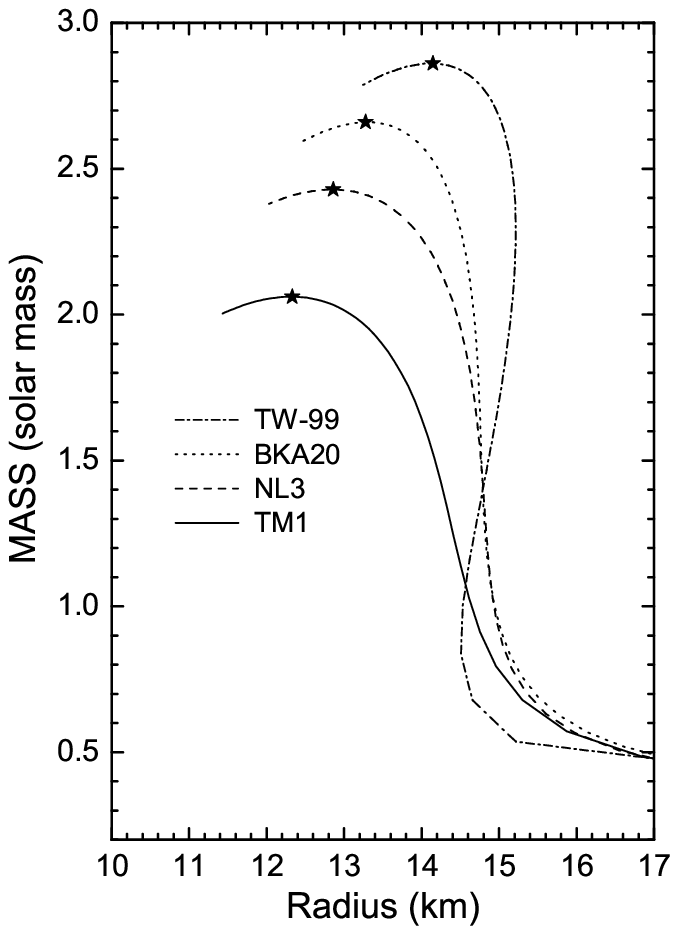}
\caption{The mass-radius relation of neutron stars. The maximum mass
is marked with an asterisk on each curve.}\label{fig NSmass}
\end{minipage}
\hfill
\begin{minipage}[t]{0.48\linewidth}
\centering
\includegraphics[width=7.cm,height=7cm]{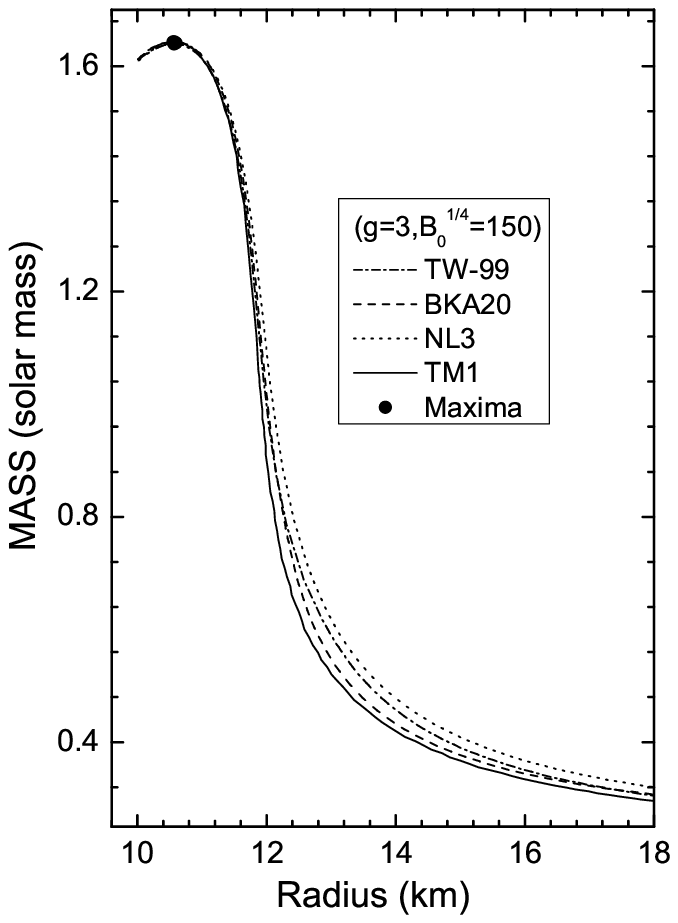}
\caption{ The mass-radius relation of hybrid stars with different
parameter sets of the nuclear matter. The maximum mass, marked with
a full dot on each cure, do not change significantly with the
nuclear parameters.} \label{fig HyS2}
\end{minipage}
\end{figure}

Before investigating the properties of the hybrid star, let's give
the mass-radius relations for compact stars composed of pure nuclear
matter. The TOV equation is solved with the nuclear matter EOS
introduced in Sec. \ref{Sec:thermo}. We take the four parameter sets
in the calculation: TM1, NL3, BKA20, TW-99. The maximum masses of
neutron stars are located in $2\sim 3$ times the solar mass
M$_\odot$ with the corresponding  radii in the range (12.3 $\sim$
14.1 km) in Fig. \ref{fig NSmass}. For harder EOS of the parameter
set TW-99, the maximum star mass approaches the value $2.86$
M$_\odot$, which is bigger than the mass $2.06$ M$_\odot$ with the
parameter set TM1. Different with the self-bound stars, the equation
of state of nuclear matter at lower density is calculated in the BPS
model.

To investigate the influence of model parameter on the mass-radius
relation of hybrid stars, we should consider different EOS of the
nuclear matter and strange quark matter. Firstly, the parameter set
($g=3$, $B_0^{1/4}=150$ MeV) is employed on the quark side, while
different parameter sets on the nuclear side are adopted in the
calculation. The corresponding mass-radius relations are plotted in
Fig. \ref{fig HyS2} where the maximum star mass are marked with full
dots. Even though the nuclear parameter is adopted for different
sets, the maximum mass is very close to the same value $1.64$
M$_\odot$. Doing calculations with other choices of the parameter
$B_0$ and $g$, we find the result that the maximum mass does not be
significantly influenced by the parameter uncertainty of the nuclear
equation of state.

\begin{figure}[ht]
\begin{minipage}[t]{0.48\linewidth}
\centering
\includegraphics[width=7.cm,height=7cm]{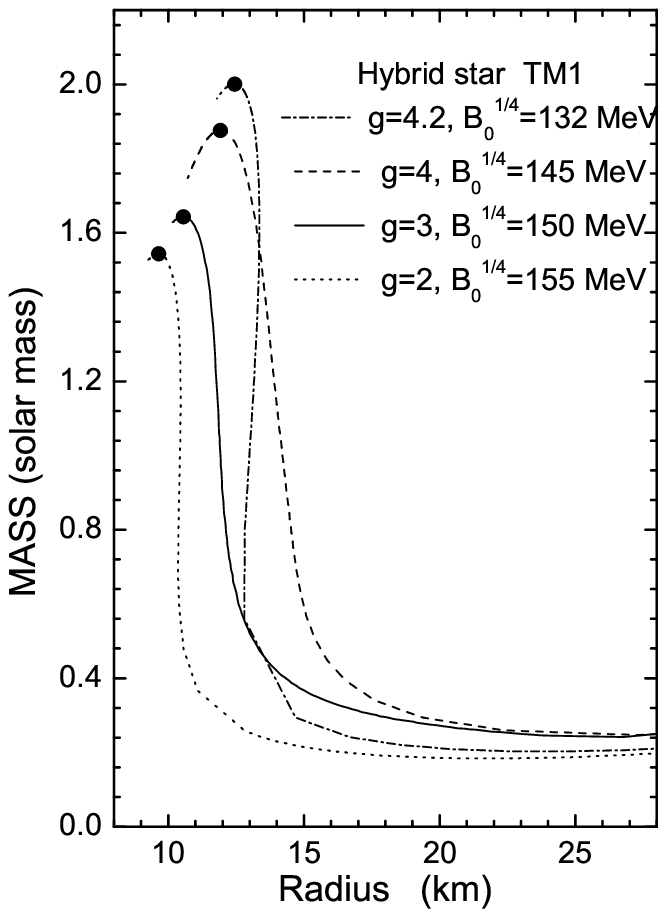}
\caption{The mass-radius relation of hybrid stars with various
parametrization of quark matter. The maximum masses are marked with
full dots.} \label{fig HyS1}
\end{minipage}
\hfill
\begin{minipage}[t]{0.48\linewidth}
\centering
\includegraphics[width=7.cm,height=7cm]{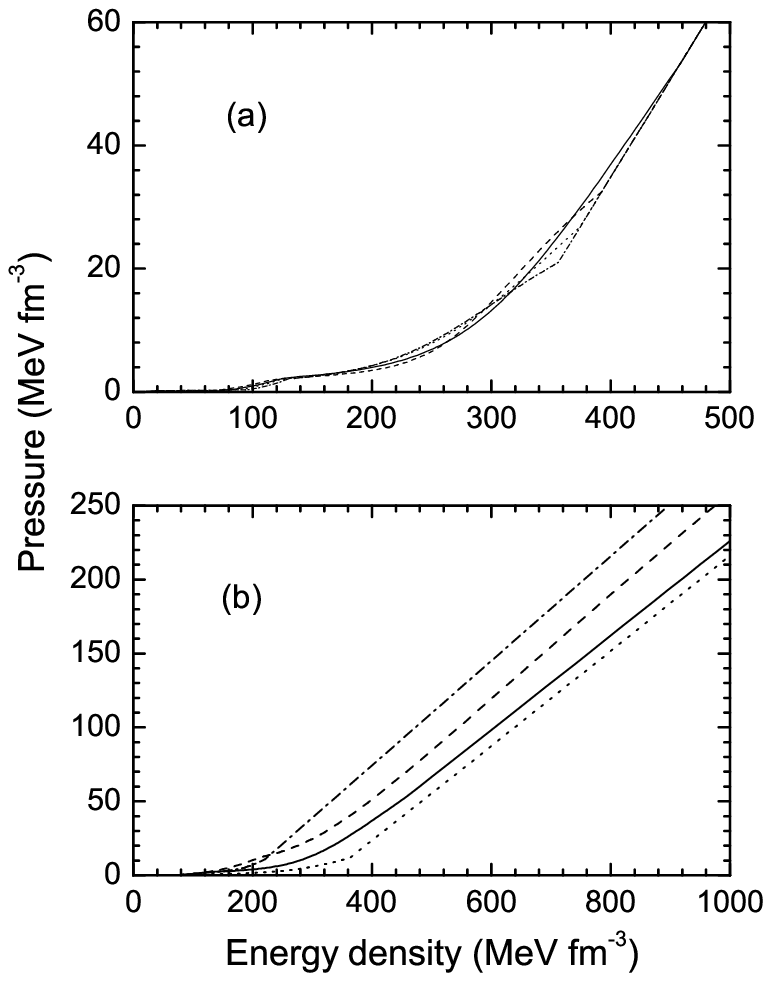}
\caption{The pressure versus the energy density for the equations of
state of hybrid stars. Figure (a) and (b) are, respectively, for
Fig.~\ref{fig HyS2} and \ref{fig HyS1}. } \label{fig eos}
\end{minipage}
\end{figure}

\begin{figure}[htb]
\centering

\includegraphics[width=7.cm,height=7cm]{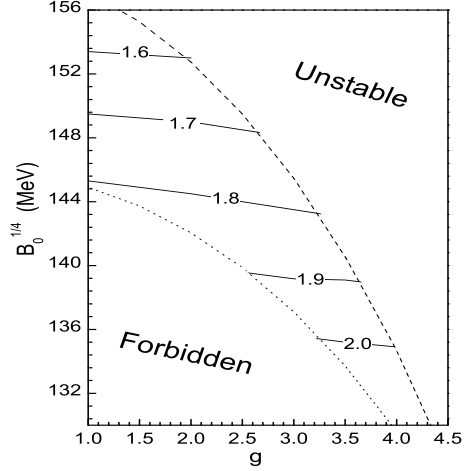}
\caption{ The contour plots of the maximum mass of hybrid stars in
the panel (B$_0^{1/4}$, $g$)  . Based on the stability window in
Fig.~\ref{fig B0window}, the maximums of hybrid stars are shown on
each line in the allowed area.} \label{fig contour}
\end{figure}

Now we fixed the TM1 parameter set for the nuclear EOS while
changing the values of $g$ and $B_0$ on the quark side.
In this case the maximum mass of hybrid stars is calculated in
Fig.~\ref{fig HyS1}. The distinction between Fig.~\ref{fig HyS2} and
\ref{fig HyS1} can be understood from the comparison of EOS. In the
upper panel of Fig.~\ref{fig eos}, we can know that the narrow range
on the energy density axis is affected by nuclear parameter sets,
while a larger range is greatly dominated by quark matter parameter
sets. For a bigger value of the maximum of hybrid stars, the
corresponding density range of the mixed phase is wider. In the
parameter range, the maximum mass is in the range $(1.5 \sim 2.04)
M_\odot$ and radius in the range ($9.64 \sim 12.4)$ km, which
comprises the pulsar PSR J1614-2230 with $1.97\pm 0.04 M_\odot$
\cite{1614}. For a completely relation between the maximum mass of
hybrid stars and the parameter set of SQM, we show the contour plots
of the maximum mass in the panel with $B_0^{1/4}$ and $g$ on the
horizontal and vertical axis in Fig.~\ref{fig contour}. According to
the stability window shown in Fig~.\ref{fig B0window}, the maximum
mass of hybrid stars has been marked on each line in the allowed
area.

\section{Summary}
\label{Sec:conls}

We have studied the hybrid stars with mixed phase of nuclear matter
and SQM. In the outer layer, the nuclear matter is described by the
relativistic mean-field theory and crust EOS by the BPS model. In
the inner core, the quark matter is investigated by the
quasiparticle model. According to the Witten-Bodmer hypothesis, we
present the stability window of quark matter in the coupling
constant $g$ versus bag constant $B_0^{1/4}$ panel. The mixed phase
exists in the range $1 \sim 3$ times the nuclear saturation density,
which is energetically favorable by comparison with pure nuclear
matter or SQM. We point that the mixed phase of hybrid stars is
located in the meta-stable or unstable regions of pure quark matter.
Otherwise the hybrid star will collapse to a strange star and no
mixed phase could exist.

We show that the maximum mass and radius of hybrid stars are
controlled by the equation of state of quark matter
\cite{Narain2006} rather than that of the nuclear model. Or more
exactly in the present framework, they are mainly controlled by the
coupling constant $g$ and bag constant $B_0$ in the quasiparticle
model. For a smaller coupling constant, the maximum mass is 1.54
M$_\odot$ with radius about 9.64 km.  The larger coupling constant
will broaden the range of the mixed phase and increase the maximum
mass of a hybrid star up to $2.0$ M$_\odot$. So the quark matter
with strong coupling can meet the constraint at high density in
compact stars \cite{klahn2006} and corroborate the ``masquerade
effect" in the previous studies \cite{alford2005}. For even high
density, the possible CFL matter, especially the magnetized CFL
matter \cite{Ferrer2011}, should be considered, which will be
investigated in our future work.

The authors would like to thank support from the National Natural
Science Foundation of China (11005071 and 11135011) and the Shanxi
Provincial Natural Science Foundation (2011011001).

\bibliographystyle{model1a-num-names}
 
\end{document}